\documentclass[universe,article,submit,moreauthors,pdftex]{mdpi} 
  
\firstpage{1} 
\pubvolume{1}
\issuenum{1}
\articlenumber{0}
\pubyear{}
\copyrightyear{2020} 
\datereceived{} 
\dateaccepted{} 
\datepublished{} 
\hreflink{}  
\preto{\abstractkeywords}{\nolinenumbers}

\pdfoutput=1

\usepackage{amsfonts,amssymb,yfonts}
\usepackage{xspace,psfrag,color}  
\usepackage{mathrsfs}
\usepackage{latexsym,pifont,textcomp} 
\usepackage[abs]{overpic} 
\usepackage{physics} 
\usepackage{lipsum}

\newcommand{\scri}{\mathscr{I}}

\def\x{{\mathbf x}}%
\psfrag{t}[][]{\Large$t$}
\psfrag{r}[][]{\Large$r$}
\psfrag{ip}[][]{$i^+$}
\psfrag{im}[][]{$i^-$}
\psfrag{i0}[][]{$i^0$}
\psfrag{scrip}[][]{$\scri^+$}
\psfrag{scrim}[][]{$\scri^-$}

\Title{Vacuum semiclassical gravity does not leave space for safe singularities}

\Author{Julio Arrechea $^{1}$\orcidA{}, Carlos Barcel\'{o} $^{1}$\orcidB{}, Valentin Boyanov $^{2}$\orcidC{} and Luis J. Garay $^{2,3,*}$\orcidD{}}
 
\address{%
$^{1}$ \quad Instituto de Astrof\'{i}sica de Andaluc\'{i}a, IAA-CSIC, Glorieta de la Astronom\'ia, 18008 Granada, Spain; arrechea@iaa.es (J.A.); carlos@iaa.es (C.B.)\\
$^{2}$ \quad Departamento de F\'{i}sica Te\'{o}rica and IPARCOS,
Universidad Complutense de Madrid, 28040 Madrid, Spain; vboyanov@ucm.es (V.B.); luisj.garay@ucm.es (L.J.G.)\\
$^{3}$ \quad Instituto de Estructura de la Materia, CSIC, Serrano 121, 28006 Madrid, Spain}
 
\corres{Correspondence: luisj.garay@ucm.es}
\abstract{General relativity predicts its own demise at singularities, but also appears
to conveniently shield itself from the catastrophic consequences of such singularities, making them safe. For instance, if strong cosmic censorship were ultimately satisfied, spacetime
singularities, although present, would not pose any practical problems to
predictability. Here we argue that under semiclassical effects the situation should be rather different: 
the potential singularities which could appear in the theory will generically affect predictability and so one will be forced to analyse whether there is a way to regularise them. For these possible regularisations, the presence and behaviour of matter during gravitational collapse and stabilisation into new structures will play a key role.
First we show that the static semiclassical counterparts to the Schwarzschild and Reissner-Nordström geometries have singularities which are no longer hidden behind horizons. Then we argue that in dynamical scenarios of formation and evaporation of black holes, we are left with only three possible outcomes which could avoid singularities and eventual predictability issues. We briefly analyse the viability of each one of them within semiclassical gravity, and discuss the expected characteristic timescales of their evolution.}
 

\begin{document} 

\section{Introduction}

A great majority of physicists would agree that classical general relativity (GR) is not completely satisfactory, if only 
for the fact that it includes situations in which it breaks down: under mild conditions, spacetime singularities are predicted to appear evolving from regular initial conditions~\cite{Penrose1965}.
There are many types of singularities in GR, and we dare say that, sociologically, the degree of dislike received by a singularity varies very much depending on the nature of the singularity.
 
The definition  of  singularity in GR is rather subtle and still a subject of study~\cite{Geroch1968,Senovilla2012}. 
The most conservative definition is {\em the existence  of incomplete 
and inextensible causal geodesics}~\cite{HawkingEllis1973}
which, however, leaves room for a more detailed characterisation~\cite{EllisSchmidt1977, Curiel2020}.  
In our discussion here we will only consider the simple curvature singularities which appear
in exact solutions modelling stellar collapse and the simplest homogeneous cosmological models. These will provide a clear view of the most prominent characteristics of singularities without the need to enter into sub-classifications.
Among the singularities that can appear in these models, relativists have a greater dislike for naked singularities than for dressed singularities; and among the latter, more for timelike and past null singularities than for spacelike and future null singularities. 
For instance, if the only singularities appearing in GR were spacelike, as the one in a Schwarzschild black hole, or future null, many researchers might accept GR as a consistent physical theory (at least up to Planckian scales).  This is because the theory would not exhibit any practical predictability problem for the future, though the world as described by spacetime would be allowed to have (possibly multiple) singular beginnings and ends. The only major problem would then be the lack of information on how the initial conditions came about; but no physical theory provides that anyway!

At least part of the physical motivation underlying the cosmic censorship hypothesis~\cite{Penrose2002} 
is precisely to show a certain benignity for classical GR. Assuming that (strong) cosmic censorship is satisfied, the only singularities allowed to appear in the theory from evolving regular initial conditions would live inside event horizons, and moreover, would have a future causal structure similar to Schwarzschild spacetime, in the sense
of only having singularities in the causal future of observers (i.e. of either spacelike or null character). For instance, although Kerr and Reissner-Nordstr\"om black holes possess timelike singularities inside their outer horizons, they also have Cauchy horizons and are thus classically unstable under perturbations, the backreaction from which is expected to transform them into
configurations more akin to Schwarzschild in the above sense~\cite{PoissonIsrael1989,PoissonIsrael1990,Ori1991,Dafermos2017}. The resulting singularities would be future-null, possibly with a spacelike sector~\cite{Brady1995,VandeMoortel2020}, and would not cause problems for predictability in the outside universe (though the possibility of a physically sensible extension beyond the null sector of the singularity is still under debate).

Another characteristic of classical GR is that the endpoint of collapse of stellar structures is usually described by the Kerr-Newman family of geometries, which are (electro-)vacuum solutions of the theory, in which a black hole region devoid of matter is enclosed by an event horizon. This idea might be already too simplistic in classical GR~\cite{MarolfOri2013}, but nonetheless the picture of matter eventually falling into the central singularity and leaving spacetime essentially empty is indeed a commonly used one. This view is reinforced by the fact that in order to analyse the behaviour of e.g. black hole binaries, one can exclusively use the vacuum Einstein equations. It is interesting to note that the static extensions of individual black hole spacetimes one often uses (see e.g. Kruskal's maximal extension of the eternal Schwarzschild solution~\cite{Kruskal1960}), are examples of Wheeler's dictum ``mass without mass''~\cite{Wheeler1955}: all their Cauchy hypersurfaces are devoid of matter!

These classical GR insights can lead us to believe that \emph{i)} the objects resulting from gravitational collapse can be understood to a large extent independently of the behaviour and nature of the matter content (provided it satisfy the appropriate energy conditions) and that \emph{ii)} problems with predictability are avoided due to the nature of the gravitational singularities. Indeed, this way of thinking is often found in the relativistic community.

Here we argue that before accepting this line of reasoning as a good guiding principle, it first has to pass a semiclassical consistency test. In fact, we will explain how adding semiclassical effects actually works against the previous reasoning, leading to the opposite conclusions.
  
\section{Vacuum energy and the semiclassical consistency test}  

Independently of whether the zero-point fluctuations of the quantum fields gravitate or not---that is, the old cosmological constant problem~\cite{Weinberg1989}---the inhomogeneities in this fluctuating sea caused by the very curvature of spacetime should gravitate.
The Renormalised Stress-Energy Tensor (RSET) of the quantum fields residing in spacetime should act as an additional source of gravity, this being the main hypothesis behind semiclassical gravity, our best developed doorway towards the realm of quantum gravity~\cite{BirrellDavies1984}. With this source, the equations governing the evolution of the spacetime geometry become

\begin{equation}\label{Eq:Einstein}
G_{\mu\nu}=8\pi \left(T_{\mu\nu}^{\rm \,class}+\expval{T_{\mu\nu}}\right),
\end{equation}

\noindent
with $T_{\mu\nu}^{\rm \,class}$ being the stress-energy tensor of the effectively classical matter (we have taken geometric units: $G=1$, $c=1$), and with the RSET $\expval{T_{\mu\nu}}$ corresponding to the vacuum contribution of all quantum fields residing on the spacetime. The crux of this theory is calculating the RSET in an arbitrary four-dimensional spacetime, which is generally not possible analytically, and is even very difficult to achieve numerically (see e.g. \cite{Anderson1993}).\par
To obtain an approximate picture of how adding semiclassical sources modifies a geometry, we usually employ symmetry-based restrictions and simplifications to the problem. For example, a standard procedure for spherically-symmetric geometries is to integrate out the angular variables, quantise the fields in the reduced 1+1 dimensional spacetime, calculate the RSET there, and apply the result to 3+1 dimensions with the Polyakov approximation (see \cite{FabbriNavarro} for a detailed discussion).\par
For a spherically-symmetric geometry with a line element

\begin{equation}\label{geo}
    ds^2=-C(u,v)du\,dv+r^2(u,v)d\Omega^2
\end{equation}

\noindent
(with $d\Omega^2$ the line element of the unit two-sphere), the non-zero components of the RSET for a massless scalar field in the Polyakov approximation are

\begin{subequations}\label{RSET}
	\begin{align}
	\expval{T_{uu}}&=\frac{\hbar f(r)}{24\pi}\left[\frac{\partial_u^2C}{C}-\frac{3}{2}\left(\frac{\partial_uC}{C}\right)^2\right],\\
	\expval{T_{vv}}&=\frac{\hbar f(r)}{24\pi}\left[\frac{\partial_v^2C}{C}-\frac{3}{2}\left(\frac{\partial_vC}{C}\right)^2\right],\\
	\expval{T_{uv}}&=\frac{\hbar f(r)}{24\pi}\left[\frac{\partial_uC\partial_vC}{C^2}-\frac{\partial_u\partial_vC}{C}\right],
	\end{align}
\end{subequations}

\noindent
where $f(r)=1/(4\pi r^2)$. The choice of vacuum state is encoded in the choice of null coordinates $\{u,v\}$ for the temporal-radial sector with which we determine the function $C(u,v)$ from the geometry \eqref{geo}, and subsequently calculate the components of the RSET \eqref{RSET}, expressed in the same coordinate system. The difference between two such choices is given by an ingoing and outgoing flux terms, as described in~\cite{Barbado2016}.

The Polyakov approximation captures the most prominent features of the RSET in different vacuum states and is widely used to determine corrections to classical spherically-symmetric geometries \cite{Parentani1994,Fabbrietal2006,Chakraborty2015}. However, it has a clear problem at $r=0$, coming from the singular $1/(4\pi r^{2})$ multiplicative factor required for $(3+1)$-dimensional conservation of the RSET. This factor can bring about spurious pathologies when solving \eqref{Eq:Einstein} at the self-consistent level, i.e. when determining the metric from both the classical and semiclassical sources. Nonetheless, the singular character of the Polyakov RSET can be remedied by deforming the function $f(r)$ into one regular at $r=0$, at the cost of breaking the conservation of the RSET. In a static situation, where $C(u,v)$ can be expressed as a function of $r$ only, conservation is recovered simply by adding non-zero angular components to the RSET, as discussed in \cite{Arrecheaetal2020}. There, a simple regularising term was added in $f(r)$ which removes the divergence at the origin while preserving the standard Polyakov approximation at large radii. Regularisation of the Polyakov RSET is necessary whenever $r=0$ and the region around it are part of the geometry under analysis, as the pathologies at the radial origin can even propagate to other regions of the spacetime. On the other hand, these aspects can sometimes be disregarded when analysing parts of static or dynamical configurations which do not explore the vicinity of $r=0$.

With this scheme it is possible to obtain an analytical approximation to the RSET which satisfies all the necessary requirements for it to be used as a source term in semiclassical gravity. Chief among these requirements is that, when calculating the RSET on top of astrophysically observable systems with weak enough gravity (planets, stars, etc.), it results in an extremely tiny contribution which can be safely neglected, as it follows from the adequacy of classical GR to describe these systems. Semiclassical contributions are negligible for objects even as dense as neutron stars. Conversely, the magnitude of the RSET on top of a solution of GR can be used as a tool to check whether a geometry would also be a consistent solution of semiclassical gravity. As we shall discuss now, this consistency is not present for the eternal Schwarzschild and Reissner-Nordstr\"om black holes.
 
\section{Schwarzschild counterpart in semiclassical gravity}  

\begin{figure}
     \centering
     \includegraphics[width=0.5\columnwidth]{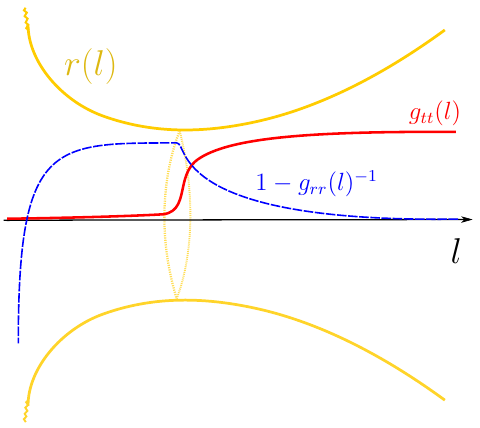}
     \caption{Pictorial representation of the semiclassical Schwarzschild geometry. The horizon is replaced by the neck of an asymmetric wormhole, which is a minimal surface for the areal radius $r(l)$, $l$ being the proper radial coordinate. Space to the right of this surface connects with the asymptotically flat region, while to the left it ends in a null singularity at finite proper distance. The red curve represents the redshift function of the geometry, which is positive everywhere but at the singularity. The blue, dashed line is the quotient between the Misner-Sharp mass and $r$, which shows how mass is distributed throughout the spacetime.}
     \label{Fig:SemiclassicalWormhole}
\end{figure}

\begin{figure}
    \centering
    \includegraphics[width=0.9\columnwidth]{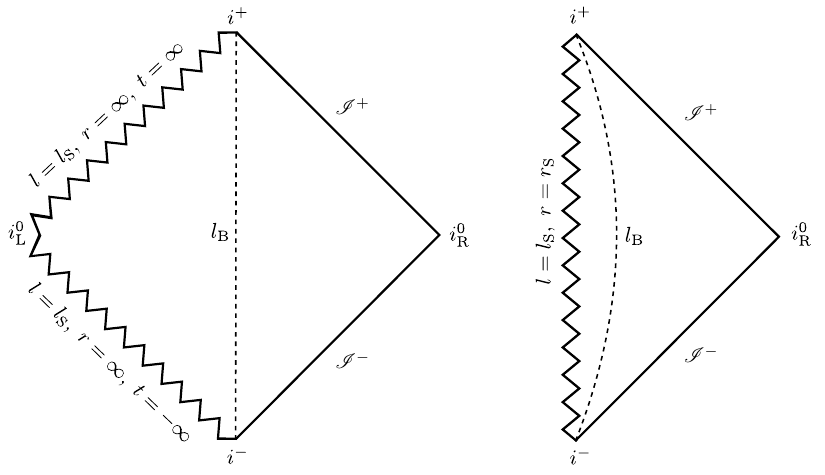}
    \caption{Left figure: Penrose diagram of the singular wormhole solution in the Polyakov approximation. The vertical dashed line $l_{\rm B}$ represents the location of the wormhole neck. To its right, the asymptotically flat portion of spacetime is depicted. The left side of the diagram shows the internal past and future null singularities, which are located at finite proper distance from the neck $l_{\rm B}-l_{\rm S}$. The point $i^{0}_{\rm L}$ is singular as well, and is reached in finite proper time by spacelike geodesics. Right figure: Penrose diagram of the singular wormhole solution in the $s$-wave approximation. The dashed curve represents the position of the wormhole neck. The singularity is timelike and located at finite radial distance from the neck, constituting a naked singularity.}
    \label{Fig:SemiclassicalWormholeCD}
\end{figure}

If one tries to generate a solution in semiclassical gravity similar to the eternal black-hole configuration, one encounters a well-known obstacle: the RSET in the Boulware vacuum state, the only genuine vacuum consistent with staticity and asymptotic flatness, diverges at the Schwarzschild horizon. 
Recently, some of the present authors performed a detailed analysis of the form of the eternal semiclassically-self-consistent counterpart of the Schwarzschild solution~\cite{Arrecheaetal2020}. Elaborating on previous works~\cite{Fabbrietal2006,HoMatsuo2018}, and with the choice of a regularised function 

\begin{equation}
    f(r)=\frac{1}{4\pi (r^{2}+\alpha l_{\rm P}^{2})},\quad \alpha>1,
\end{equation}

\noindent
with $l_{\rm P}=\sqrt{\hbar/12\pi}$, the authors found that the semiclassical solution with a positive asymptotic mass turns out to have a wormhole neck (see figure~\ref{Fig:SemiclassicalWormhole}) instead of a Schwarzschild horizon (left diagram in figure~\ref{Fig:SemiclassicalWormholeCD} shows the conformal diagram associated with these wormhole geometries). The line element around this neck, written in terms of the proper radial coordinate $l$ (defined as the proper length in the radial direction in sections of $t=\text{const.}$), is

\begin{equation}
    ds^{2}\simeq-\left[\sqrt{k_{0}k_{1}}\left(l-l_{\rm B}\right)+\phi_{\rm B}\right]dt^{2}+dl^{2}+\left[\frac{k_{1}}{4}\left(l-l_{\rm B}\right)^{2}+r_{\rm B}^{2}\right]^{2}d\Omega^{2},
\end{equation}

\noindent
where $k_{0,1}$ are positive constants that depend on $\alpha, l_{\rm P}$, and the neck radius $r_{\rm B}$, and $l_{\rm B}$ is the value of the proper radial coordinate $l$ at the neck.
The parameter $r_{\rm B}$, as well as the redshift constant $\phi_{\rm B}$ at the neck, depend on the Arnowitt-Deser-Misner mass, from which they can be obtained by solving \eqref{Eq:Einstein} numerically starting from the asymptotically flat region. As the spacetime is integrated inwards, a cloud of negative mass, coming from the effective semiclassical fluid described by the RSET, gradually modifies the geometry until, just above the Schwarzschild radius of the geometry (corresponding to the asymptotic mass), the wormhole neck is reached.
On the other side of the neck, there is a runaway decrease of the mass as $r$ increases, generating a null singularity at finite geodesic distance (for all three types of geodesics). This singularity is approached as $r\to\infty$, and the metric around it is (in Schwarzschild coordinates) approximately

\begin{equation}
    ds^{2}\simeq\left(\frac{r}{l_{\rm P}}\right)^{3-4\alpha}e^{-2r^{2}/l_{\rm P}^{2}}\left[-\frac{a_{0}}{b_{0}}\left(\frac{l_{\rm P}}{r}\right)^{2}dt^{2}+dr^{2}\right]+r^{2}d\Omega^{2},
\end{equation}

\noindent
with $a_{0},b_{0}$ being dimensionless constants. 
The structure of wormhole solutions with a mass much greater than the Planck mass is qualitatively the same independently of the particular choice of regulator parameter $\alpha$; in fact, even different approximations for the RSET~\cite{Fabbrietal2006,Arrecheaetal2020,HoMatsuo2018} lead to equivalent solutions. Only Planck-sized solutions differ significantly from one another depending on these choices, but these are discarded as they fall outside the range of validity of the semiclassical approximation itself.

Thus, essentially, semiclassical corrections in the Polyakov approximation push the future Schwarzschild horizon of classical GR (where the redshift function goes to zero) towards an internal and singular asymptotic region inside a wormhole. If one considers a more refined approximation to the RSET, such as the $s$-wave analysis in \cite{Fabbrietal2006} (this approximation incorporates the effect of back-scattering of the modes in the $s$-wave sector), then the curvature singularity inside the wormhole appears before an asymptotic region is reached and before the redshift function goes to zero (the conformal diagram of this spacetime is represented in the right diagram in figure \ref{Fig:SemiclassicalWormholeCD}). The self-consistent vacuum semiclassical solution develops a proper naked singularity.
Thus, it appears that the ``mass-without-mass" characteristic of Schwarzschild black holes is not preserved in semiclassical gravity.

These features seem to be a common result whenever one searches for the semiclassical corrections to an outer black horizon sourced by a RSET with fields in their Boulware vacuum state. Another scenario that exemplifies this (and actually contains the corrected Schwarzschild geometry described above as a particular subcase) is the semiclassical counterpart of the Reissner-Nordström geometry, which some of the present authors have obtained and analysed \cite{Arrechea2021}. The introduction of electromagnetic charge allows us to study the semiclassical corrections to a broader family of geometries with horizons. For the more physical example of a charged black hole which satisfies to $M\gg Q$ and $M\gg M_{\rm P}$, these corrections lead to wormhole geometries qualitatively similar to the previously described ones, with $Q=0$. A wormhole neck appears essentially at the location where the outer horizon of the classical solution would have appeared. Thus, there is no trace left of the inner horizon present in the classical solutions. In our Polyakov approximation the singularity inside the wormhole is null but, as stated above, including back-scattering effects would make it timelike and thus a proper naked singularity.  
By increasing the charge-to-mass ratio of the geometry the authors found
a separatrix solution reminiscent of the extremal black hole. This solution preserves a surface which appears to be a horizon, but which turns out to be a so-called non-scalar curvature singularity \cite{Arrechea2021}. We do not expect this last feature to change when considering more refined approximations to the RSET. Finally, when the charge surpasses a critical limit we find solutions with plain naked singularities (similar to those in the super-extremal regime in classical GR). The strength of the singularity is determined by a combination of quantum and classical factors whose proportion depends on the magnitude of the regulator parameter $\alpha$.

Looking at these results, we can safely say that there is a significant difference between semiclassical gravity and classical GR. Let us explain what we mean exactly (for simplicity, restricting the scope of our discussion to spherically-symmetric configurations). In classical GR the evolution of an initially dispersed cloud of material undergoing gravitational collapse leads to the formation of a black hole structure; at the same time the theory contains eternal black hole solutions (vacuum solutions, i.e. the Schwarzschild family of solutions and Kruskal's maximal extension thereof \cite{Kruskal1960}) which can be used to understand many of the characteristics of their more physical relatives. The future part of the causal structure of these two spacetimes (i.e. the black hole formed by collapse and the eternal black hole) are very similar in the sense of having a static horizon and being essentially devoid of matter. The structure of the trapped region is also independent of the matter content which, if present, collapses quickly toward the singularity. Although the extension of the eternal black hole version has two asymptotic regions, they do not causally affect each other. For these reasons the eternal solutions can be used as simpler proxies to understand black holes formed by collapse.   

Now, at this stage we have not addressed what the outcome of a semiclassically self-consistent collapse may be (which we leave for the next section). However, the above results strongly suggest that regardless of the outcome of collapse, the static vacuum semiclassical solutions would not serve as good models for understanding the future part of the resulting spacetime. One possible interpretation of this result is that the semiclassical treatment is making the second asymptotic region of the eternal black hole into a singularity, which is now in causal contact with the relevant asymptotic region, becoming a naked singularity (it appears that to recover the asymptotic flatness of the second region, one would need to make the wormhole completely symmetric by adding some classical matter to the system). In the following we are going to argue that in order to avoid problems of predictability in any realistic process of collapse, it is crucial to understand the role of classical matter, particularly how its effect on the geometry influences the quantum vacuum of fields, potentially making the resulting semiclassically self-consistent solutions completely different.

\section{Semiclassical collapse and subsequent evolution}

As is well known, when a collapse process progresses to the point of forming a trapping horizon (particularly, when this happens in the standard rapid fashion), the quantum fields evolve self-consistently in such a way that when the horizon is about to form, these fields are no longer in a Boulware state but rather in an Unruh state.
(By trapping horizon we technically mean a momentaneous or apparent horizon; as our discussion here is restricted to spherical symmetry there is no problem in uniquely identifying this surface).
Quantum fields in this state have mild energy contributions around the horizon which are inconsequential to short-term dynamics. Indeed, if we are not in a quasi-stationary situation, semiclassical effects are proven to be negligible around an outer apparent horizon, except for the ignition of a tenuous evaporation process~(see e.g. \cite{Bardeen1981,Boyanovetal2019}).  
While the outer horizon starts to slowly shrink inwards, matter inside the horizon continues its semiclassical evolution. This evolution is at least initially a continuation of the collapse process, meaning that there is a transient period in which the geometry is indeed similar to that of a  black hole with a trapping horizon. However, this only delays facing the problem of the final fate of the collapsing matter and its potential consequences.

Assuming that spacetime admits at least an effective classical geometric description, there are essentially four possibilities for its remaining future development. We will now look at each of them in relation to the problems of singularities and predictability.

\subsection{Evaporation {\em \`a la} Hawking:}

\begin{figure}
     \centering
     \includegraphics[width=0.3\columnwidth]{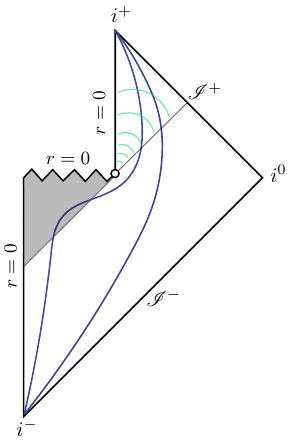}
     \caption{Causal diagram of a Hawking-like evaporating black hole. The shaded portion corresponds to the interior of the event horizon, which exists up to the end point of the evaporation process. The blue curves joining $i^-$ and $i^+$ qualitatively represent \mbox{$r=$ const.} surfaces. The green wave fronts represent the causal effect of the thunderbolt on future events.}
     \label{Fig:HawkingEvaporationDiagram}
\end{figure}

The process is described in the causal diagram of figure~\ref{Fig:HawkingEvaporationDiagram}. This description assumes that a singularity is indeed formed during the collapse.
The black hole initially has a Schwarzschild-like structure consisting of a spacelike singularity covered by a trapped region. Evaporation of the trapped region occurs slowly, but eventually the outer horizon meets the singularity, resulting in the so-called \textit{thunderbolt} event. The time it takes for this to occur as obtained in \cite{Barceloetal2020} (compatible with Hawking's original calculation \cite{BHexpl}) through backreaction of the RSET in the Polyakov approximation is
\begin{equation}
v_{\rm H}=\frac{64}{3} \frac{M^3}{l_{\rm P}^2}\simeq \left(\frac{M}{M_\odot}\right)^3 10^{73}\,\text{s},
\end{equation}
\noindent
where $M$ is the initial black hole mass.

While the trapping horizon is far (in radial distance) from the singular region, we are in a situation similar to classical GR. However, from the thunderbolt event onward, predictability is put in jeopardy.
On the one hand, there is the problem of information loss associated 
with the evaporation process~\cite{Hawking1976,Wald1984}. On the other hand, it becomes necessary to know the complex features of the thunderbolt in order to predict any future events from that point on. Since this event is causally connected with future infinities, the slowness of the Hawking evaporation only delays the appearance of potential predictability difficulties in this scenario.

However, other scenarios which alleviate these issues are in principle possible, and they strongly depend on the behaviour of matter and semiclassical gravity in dynamical regimes close to the formation of a classical singularity. It may be the case that only a complete theory of quantum gravity can solve this issue completely, but here we want to exhaust the possibilities which can be described in the semiclassical framework. 

For instance, let us assume that the semiclassical evolution is such that it results in a completely regular spacetime.
In spherical symmetry, once a trapping horizon is formed, the different possibilities for the subsequent evolution resulting in geodesically complete spacetimes have been classified in~\cite{Carballo-Rubio2019}. 
Within this catalogue there are several configurations which we are going to discard because they exhibit some unphysical features. Specifically, the catalogue contains:
\begin{itemize}
\item
Regular black-hole configurations~\cite{Bardeen1968}. These configurations contain Cauchy horizons which would be strongly unstable at the semiclassical level~\cite{FrolovZelnikov2017}. These geometries appear to be semiclassically inconsistent, unless they are just a good approximation to the evolving geometry for a brief transient period. Any semiclassically self-consistent approach should avoid their strict formation.

\item
Wormhole geometries. As we are considering only scenarios of deterministic evolution that start from initial data associated with a regular collapsing star. Thus, we will exclude geometries which involve a topology change (the above case of strict formation of regular black holes would also involve topology change of the putative Cauchy surfaces).

\item
Geometries with trapped regions touching infinity. We do not have any strong argument against these configurations. However, they can be considered (infinitely) stretched versions of those with trapping regions of finite size. We will not directly contemplate this possibility in the following.

\end{itemize}
After discarding these configurations we are left with essentially three possibilities, which we will describe in the following.

\subsection{Regular black hole evaporating inwards from the outside:}

\begin{figure}
     \centering
     \includegraphics[width=0.3\columnwidth]{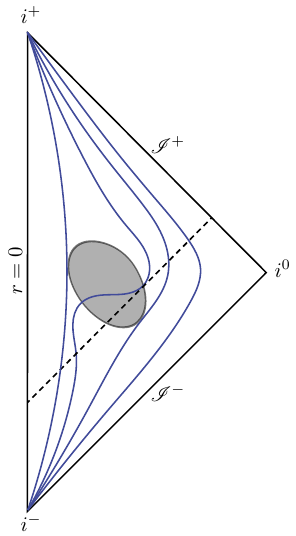}
     \caption{Causal diagram of an evaporating regular black hole. The shaded region represents the black hole trapped region. The blue curves joining $i^-$ and $i^+$ qualitatively represent \mbox{$r=$ const.} surfaces.}
     \label{Fig:EvaporatingRegularBH}
\end{figure}

These types of configurations have an internal (inner trapped) apparent horizon which never reaches the origin (see figure~\ref{Fig:EvaporatingRegularBH}) \cite{RomanBergmann1983}. Semiclassically, this internal horizon should not lead to the formation of a Cauchy horizon, due to the evaporation of the trapped region. Rather, it either remains at a finite radius until the outer horizon meets it, or has an evolution of its own which must be analysed. 
The former of these two possibilities represents the smallest deviation from Hawking's paradigm, having a long-lived external horizon, but avoiding the formation and subsequent ``evaporation" of a singularity. Given the regularity of the resulting geometry, this possibility need not lead to any information loss. From an astrophysical perspective, such an object would be indistinguishable from a standard classical black hole. Its lifetime would be roughly the same as in the Hawking scenario, much larger than the age of the universe. However, we note that this possibility is unlikely, as long-lived static inner horizons are highly unstable semiclassically~\cite{FrolovZelnikov2017,Carballo-Rubio2019}.

\subsection{Regular black hole evaporating outwards from the inside:}

A recent analysis by some of the present authors \cite{Barceloetal2020} has shown that inner horizons have a semiclassical initial tendency to evolve outwards, i.e. to make the trapped region disappear from the inside outwards. Extrapolating from this initial tendency, the time it would take for the inner horizon to reach the outer one and make the trapped region disappear is estimated to be

\begin{equation}
v_{\rm evap}\simeq \frac{1}{\kappa_{1,0}+(2r_{\rm i,0})^{-1}}\log\frac{M}{M_{\rm P}}\lesssim\frac{M}{M_\odot}10^{-4}\,\text{s},
\end{equation}

\noindent
where $r_{\rm i,0}$ and $\kappa_{1,0}$ are the initial position and surface gravity of the inner horizon. For the upper bound on the right we assume $\kappa_{1,0}> 1/M$, which is the scale of the surface gravity at the outer horizon. The causal diagram in this case is qualitatively the same as before (figure~\ref{Fig:EvaporatingRegularBH}). The crucial difference is the extremely different lifetime of the trapped region as seen by external observers.

This suggests that the most likely possibility from a semiclassical point of view is one in which the trapped region is short lived. Additionally, both in this and in the previous scenario, the behaviour of matter is crucial in the prevention of singularity formation and can strongly influence the final stages of evaporation. Predictability is once again safe because of the regularity of the entire construction, which prevents thunderbolt-like events, and not because of the presence of only spacelike singularities. After this process is complete, what comes into causal contact with future events is a central core which had always remained regular.

For an astronomical observer, the mass contained in the object may not change much during this process, as the total mass evaporated through the outer horizon would be very small in the short period it takes for the trapped region to disappear.
What form would the resulting object then acquire? One possibility is that it may reenter a collapsing phase with fast enough dynamics for a (transient) trapped region to once again form. Another is that it may enter a dynamical regime through which it stabilises into a new form of static stellar equilibrium. To understand how this latter possibility may be realised, we first note that taking the final horizonless static configuration as a background, the vacuum state of a quantum field would quickly relax to the Boulware state. Conversely, the energy content of this state may be what brings about equilibrium in the first place, if the initial dynamics are close enough to staticity and the surface of the object is close enough to its gravitational radius. In other words, when the dynamical process is no longer an almost free-falling collapse from an initial object whose surface is very far from its gravitational radius, hints of the energy content of the Boulware vacuum can be revealed \cite{Boyanovetal2019}. Then, the available evidence strongly suggests that the new states of equilibrium should be ultracompact stars with no horizons (for a more detailed description of how a transient state might lead to equilibrium see~\cite{Barceloetal2016u}; for a detailed analysis of how semiclassical gravity could lead to ultracompact configurations see~\cite{Carballo-Rubio2018} and \cite{Arrechea2021b}).

\subsection{Time-symmetric bounce:} 

\begin{figure}
     \centering
     \includegraphics[width=0.3\columnwidth]{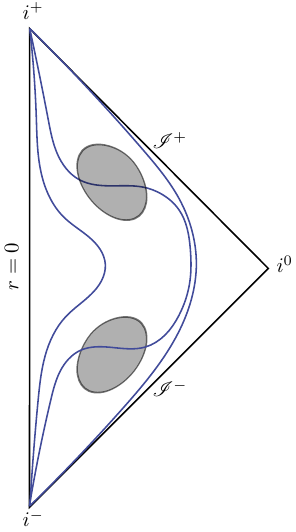}
     \caption{Causal diagram of a bouncing geometry. The lower shaded region is a black-hole-like trapped region, while the upper one is a white-hole-like trapped region. The blue curves  joining $i^-$ and $i^+$  qualitatively represent \mbox{$r=$ const.} surfaces.}
     \label{Fig:TS-BounceDiagram}
\end{figure}

In this scenario, gravitational collapse is followed by an (approximately) time-symmetric bounce \cite{Barceloetal2011,Barceloetal2015,HaggardRovelli2015} (see figure~\ref{Fig:TS-BounceDiagram}). There are two trapped regions, the bottom one is outer-trapped and the upper one inner-trapped. Again, any long-lived white horizon should be unstable, if for no other reason than just semiclassical consistency~\cite{Barceloetal2016,Barceloetal2020}. 
Assuming that the trapped regions are short lived, one must consider this bouncing geometry as just describing one cycle within a transient period consisting of oscillations between black-hole-like and white-hole-like configurations. After this transient period, dissipation would cause the system to search for a new (horizonless) state of static equilibrium. Much like in the previous case, the stabilisation of these configurations would be closely related to the energy content present in the Boulware vacuum. The new states of equilibrium should then be ultracompact stars with a surface slightly above the gravitational radius. Once again, in this scenario the behaviour of matter is crucial for understanding both the transient regime and the static equilibrium states.

As we have described above, self-consistent configurations in the Boulware vacuum have naked singularities and, being wormholes, are not a sensible result for the evolution of an initially regular star. As we thoroughly discuss in~\cite{Arrechea2021b}, the introduction of some classical matter could result in regular ultracompact solutions (strictly speaking, that work proves that the regularised Polyakov approximation is sufficient to produce quasi-strictly-regular ultracompact configurations). In this scenario predictability is once again safe because of the regularity of the entire construction, not because of the presence of only spacelike singularities.

Before ending this section, let us briefly make some speculative comments on which might be the semiclassical version of the classical process by which two
gravitational waves collide to form a black hole. In an impressive mathematical treatise Christodoulou \cite{Christodoulou2008} has proved that the collision of two realistic gravitational waves (i.e. two localised wave fronts) can lead to the formation of closed trapped surfaces and so to black holes. In classical GR, this provides another argument in favour of the relevance of the vacuum Einstein equations. In fact, this can be argued to be an even clearer example of ``mass without mass''. However, we have reasons to expect that this situation would be strongly changed when quantum effects are taken into account. Semiclassical physics once again gives the first indication that this would be the case, as vacuum polarisation of quantum fields, however small, takes us away from the classical GR vacuum. Particularly, in the context of the process of black-hole to white-hole transitions explored in papers like~\cite{Barceloetal2011,HaggardRovelli2015,Barceloetal2015}, one may expect that under quantum effects the collapse of gravitational radiation may lead to a complete dissipation of the initially concentrating energy (although we cannot completely discard the possibility that some gravitational geon~\cite{Anderson1997} is formed after the collapse and relaxation phase).
This leads to the appealing speculation that the only stable stellar-like configurations permitted by semiclassical gravity require incorporating some matter (as opposed to radiation) component. Interestingly enough, this would bring semiclassical gravity closer than classical GR to the initial Machian character Einstein was seeking in his theory.

\section{Conclusions} 

All in all, adding semiclassical effects to GR completely changes the line of reasoning we started from: \emph{i)} Understanding the behaviour of matter is essential for predicting the actual result of gravitational collapse; as opposed to what happens in classical GR, vacuum semiclassical gravity by itself is not as good a proxy to understand the final outcome of collapse. And \emph{ii)} if predictability is ultimately safe under semiclassical collapse, it would not be because the process tends to form Schwarzschild-like configurations, with their dressed end-of-spacetime singularities (this would be semiclassically inconsistent); it would be because the behaviour of matter regularises the singular regions, and only then puts them in causal contact with the external asymptotic future.

Although the static Boulware vacuum does not enter the analysis of the initial collapsing phase of a star, it is important to realise that, in combination with classical matter in certain dynamical regimes (quite possibly reached through semiclassical dynamics), it can be behind the realisation of some new forms of equilibrium in the form of ultracompact stars. It is still an open question whether the objects that in the astrophysical parlance are referred to as black holes, might instead be horizonless ultracompact objects.

\vspace{6pt} 

\authorcontributions{Conceptualisation, J.A, C.B., V.B., and L.J.G.; formal analysis, J.A, C.B., V.B., and L.J.G.; investigation, J.A, C.B., V.B., and L.J.G.; writing—original draft, J.A, C.B., V.B., and L.J.G.; prepara-
tion, J.A, C.B., V.B., and L.J.G.; writing—review and editing, J.A, C.B., V.B., and L.J.G.; funding acquisition,
C.B. and L.J.G.. All authors have read and agreed to the published version of the manuscript.}

\funding{This research was funded by the Spanish Government through the projects FIS2017-86497-
C2-1-P, FIS2017-86497-C2-2-P (with FEDER contribution) and FIS2016-78859-P (AEI/FEDER,UE),
and by the Junta de Andalucía through the project FQM219. C.B. and J.A. acknowledge financial support
from the State Agency for Research of the Spanish MCIU through the “Center of Excellence Severo
Ochoa” award to the Instituto de Astrofísica de Andalucía (SEV-2017-0709). V.B. is funded by the Spanish Government fellowship FPU17/04471.}

\institutionalreview{Not applicable.}

\informedconsent{Not applicable.}

\dataavailability{Not applicable.}

\conflictsofinterest{The authors declare no conflict of interest.} 
\end{paracol}
\reftitle{References}



\end{document}